\documentclass{jfm}
\usepackage{graphicx}
\usepackage{epstopdf, epsfig}
\usepackage{caption}
\usepackage{subcaption}
\usepackage{amsmath}
\usepackage{amssymb}
\usepackage{tikz}
\usetikzlibrary{shapes}
\definecolor{blue}{RGB}{0,112,192}
\definecolor{lightblue}{RGB}{0,176,240}
\definecolor{green}{RGB}{0,176,80}
\definecolor{yellow}{RGB}{255,255,0}
\definecolor{orange}{RGB}{255,192,0}
\definecolor{red}{RGB}{255,0,0}
\definecolor{darkred}{RGB}{118,0,0}
\definecolor{purple}{RGB}{208,0,154}

\shorttitle{Axisymmetric column collapses of bi-frictional granular mixtures}
\shortauthor{T. Man, Z. Zhang, ,H. E. Huppert, S. A. Galindo-Torres}

\title{Axisymmetric column collapses of bi-frictional granular mixtures}

\author{Teng Man\aff{1},
  Zaohui Zhang\aff{1,2},
  Herbert E. Huppert\aff{3}
 \and Sergio A. Galindo-Torres\aff{1}
 \corresp{\email{s.torres@westlake.edu.cn}}
}

\affiliation{\aff{1}Key Laboratory of Coastal Environment and Resources of Zhejiang Province (KLaCER), School of Engineering, Westlake University, 600 Dunyu Rd, Hangzhou, Zhejiang 310024, China
\aff{2}College of Environmental and Resource Sciences, Zhejiang University, 866 Yuhangtang Rd, Hangzhou 310058, China
\aff{3}Institute of Theoretical Geophysics, King's College, University of Cambridge, King's Parade, Cambridge CB2 1ST, UK}

\begin{document}

\maketitle

\begin{abstract}
The behavior of granular column collapses is associated with the dynamics of geohazards, such as debris flows, landslides, and pyroclastic flows, yet its underlying physics is still not well understood. In this paper, we explore granular column collapses using the spheropolyhedral discrete element method (DEM), where the system contains two types of particles with different frictional properties. We impose three different mixing ratios and multiple different particle frictional coefficients, which lead to different run-out distances and deposition heights. Based on our previous work and a simple mixture theory, we propose a new effective initial aspect ratio for the bi-frictional granular mixture, which helps unify the description of the relative run-out distances. We analyze the kinematics of bi-frictional granular column collapses and find that deviations from classical power-law scaling in both the dimensionless terminal time and the dimensionless time when the system reaches the maximum kinetic energy may result from differences in the initial solid fraction and initial structures. To clarify the influence of initial states, we further decrease the initial solid fraction of granular column collapses, and propose a trial function to quantitatively describe its influence.  Due to the utilization of a simple mixture theory of contact occurrence probability, this study can be associated with the friction-dependent rheology of granular systems and friction-induced granular segregations, and further generalized into applications with multiple species of particles in various natural and engineering mixtures.
\end{abstract}


\section{Introduction}
\label{sec:intro}
Granular materials are ubiquitous in natural and engineering systems, such as debris flows, landslides, fresh concrete, and fissured rocks. Understanding the constitutive behavior of granular media is significant for solving problems in at least civil engineering, chemical engineering, and pharmaceutical engineering. Progress has been made since the proposal of Bagnold rheology \citep{bagnold1954experiments}, where both normal and shear stresses are proportional to $f(\phi_s)\rho_p\dot{\gamma}^{2}d^2$, and the $\mu(I)$ rheology \citep{midi2004dense,jop2006constitutive,pouliquen2006flow}, where the effective frictional coefficient, $\mu = \tau/\sigma_n$, can be expressed as a function of the inertial number, $I = \dot{\gamma}d/\sqrt{\sigma_n/\rho_p}$ [where $f(\phi_s)$ is a function of the solid fraction, $\phi_s$, $\rho_p$ is the particle density, $\dot{\gamma}$ is the shear rate, $d$ is the average particle size, $\tau$ is the shear stress, and $\sigma_n$ is the pressure].

With the successful characterization of dry granular systems in steady-states, granular column collapses were proposed to investigate the transient behavior and to relate granular flows to natural geophysical flows, such as pyroclastic flows and landslides \citep{Roche2002ExperimentsOD,lacaze2009axisymmetric}. \citet{lube2004axisymmetric} and \citet{lajeunesse2005granular} tested the collapse morphology and kinematics of dry granular column collapses and concluded a power-law relationship between the initial aspect ratio, $\alpha = H_i/R_i$, and the relative run-out distance, $\mathcal{R}=(R_{\infty} - R_i)/R_i$, where $H_i$ is the initial height of the column, $R_i$ is the initial column radius, and $R_{\infty}$ is the final run-out radius after the column collapse. Based on the $\mathcal{R}(\alpha)$ relationship, a critical aspect ratio, $\alpha_c$, was observed to divided granular column collapses into two regimes that (1) when $\alpha < \alpha_c$, $\mathcal{R}$ is approximately proportional to $\alpha$, and (2) when $\alpha > \alpha_c$, $\mathcal{R}$ approximately scales with $\alpha^{0.5}$ \citep{lube2004axisymmetric,lube2005collapses,thompson2007granular}. \citet{zenit2005computer} performed discrete element method (DEM) simulations on two dimensional (2D) granular column collapses, and confirmed that the shape of the final deposition was mainly determined by the initial aspect ratio. \citet{staron2005study,staron2007spreading} further investigated 2D granular column collapses with DEM, and found that the inter-particle frictional coefficient played an important role in the run-out distance, but did not quantify such frictional effects. Previous research also studied the complexity of granular column collapses when the system was subjected to different realistic conditions, such as particle size polydispersity \citep{Cabrera2019granular,Martinez2022segregation}, fluid saturation or immersion \citep{rondon2011granular,fern2017granular,Bougouin2019collapse}, complex particle shapes \citep{Zhang2018influence}, and erodible bottoms \citep{Wu2021collapse}. However, no matter how complex the granular system was, the inter-particle friction was often set constant and unique.

To account for the influence of both the inter-particle friction and boundary friction, \citet{man2021deposition} proposed an effective aspect ratio, 
\begin{equation} \label{eq-alpha-eff}
    \alpha_{\rm eff} = \alpha\sqrt{1/(\mu_w + \beta\mu_p)}\ ,
\end{equation}
based on a dimensional analysis, where $\mu_w$ is the frictional coefficient between particles and the bottom plate, $\mu_p$ is the frictional coefficient between contacting particle pairs, and $\beta$ is a fitting parameter, and later linked $\alpha_{\rm eff}$ to the ratio between inertial effect and frictional resistance existing in granular systems during the collapses. Similar to studies of \citet{warnett2014scalings,Cabrera2019granular}, \citet{man2021finitesize,man2022influence} also observed the size effect of the granular column collapses but further related the size effect to finite-size scaling (FSS) and characterized the influence of cross-section shapes using the FSS analysis, so that the size effect of granular column collapses can be quantified as
\begin{equation} \label{eq:sizeeffect}
    \mathcal{R} = \left({R_i}/{d}\right)^{-\beta_1/\nu}\mathcal{F}_r\left[(\alpha_{\rm{eff}}-\alpha_{c\infty})\left({R_i}/{d}\right)^{1/\nu}\right]\ ,
\end{equation}
where $\mathcal{F}_r[\cdot]$ is a scaling function, scaling parameters $\nu = 1.39\pm 0.14$ and $\beta_1 = 0.28\pm 0.04$ are obtained to best collapse all the data, and $\alpha_{c\infty}$ is the transitional effective aspect ratio when the system size goes to infinity. The influence of friction effects on granular column collapses resembled the friction-dependent rheology we proposed earlier \citep{man2023friction}, where the frictional rheology depended on a frictional number, $\mathcal{M}$, which was also a ratio between inertial effects and frictional resistance.

However, no granular assembly in nature constitutes only one species of grains. A granular mixture may involve particles with different degrees of roughness and different angularities, which result in different inter-particle frictional coefficients. We have confirmed the influence of frictional coefficient in our previous studies \citep{man2021deposition,man2023friction}, but have not yet explored the condition when a granular system contains particles with different friction properties. In this paper, we aim to address this issue by introducing a bi-frictional granular mixture, where the system includes two species of particles, Grain\#1 and Grain\#2, with different inter-particle frictional coefficients, to investigate the mixing effect associated with granular column collapses using DEM. This paper is organized as follows. In Sect.\ref{sec:exp}, we provide a set of experimental examples to show the influence of mixing particles with different frictional coefficients. In Sect.\ref{sec:simu}, we introduce both the DEM model and the simulation setup, and define essential parameters. We then elaborate the simulation results and provide several discussions in Sect.\ref{sec:res_discuss} to illustrate and quantify the mixing effect of bi-frictional systems, before providing some concluding remarks in Sect.\ref{sec:conclu}.

\section{Experimental setup and results}
\label{sec:exp}
To experimentally verify the friction-dependency of granular column collapses, we acquired two different type of particles. Grain\#1 is irregular-shaped glass particles with diameter, $d_1$, ranging from 1 mm to 3 mm (diameter range is obtained from sieve tests) and particle density, $\rho_1$, being approximately equal to 2.678 g/cm$^3$. Grain\#2 is river sand particles with diameter, $d_2$, also ranging from 1 mm to 3 mm and particle density, $\rho_2\approx$2.664 g/cm$^3$. We used a transparent plastic cylindrical tube to form the initial granular column, and the initial radius of tested granular columns was 23 mm. We varied the amount of granular materials poured into the cylindrical tube to achieve different initial packing height ranging from $\approx 3$ mm to $\approx 146$ mm and resulting in the initial aspect ratio, $\alpha$, ranging from 0.13 to 6.34. After placing particles into the cylindrical tube, we measured the initial height of the granular packing, $H_i$. Particles were dropped in from the top of the tube so that the initial condition resembled a randomly loose packing of the granular system. Then, the tube was manually lifted to release all the particles to form a granular pile. We measured the final radius of the sand pile in eight different directions and took their average as the final run-out distance, $R_{\infty}$. Then, the relationship between the initial aspect ratio, $\alpha$, and the normalized run-out distance, $\mathcal{R}$, can be obtained accordingly.

\begin{figure}
  \centerline{\includegraphics[scale=0.4]{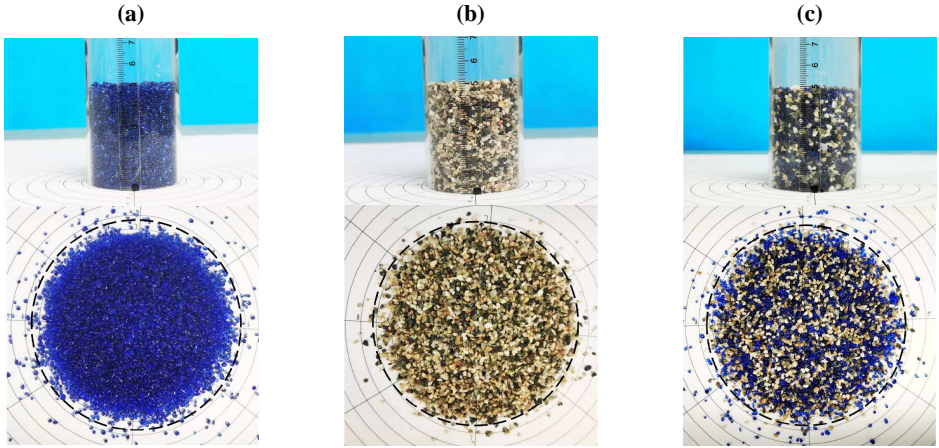}}
  \caption{Initial configurations and final depositions of granular column collapses of (a) 100\% irregular-shaped glass particles, (b) 100\% sand particles, and (c) 50\% glass particles + 50\% sand particles. The cylindrical radius is 23 mm and the initial height in the figure is approximately 50 mm.}
\label{fig:exp_setup}
\end{figure}

Three sets of experiments were performed in this study: (1) 100\% glass particles; (2) 100\% sand particles; and (3) 50\% glass + 50\% sand particles. In Figure \ref{fig:exp_setup}(a-b), we show the initial state and the final deposition of a simulation of 100\% glass particles. We can see that the glass particles have irregular shapes that resemble the shape of river sand. The initial height was 50 mm, and the final run-out distance was approximately 78 mm. In Figure \ref{fig:exp_setup}(c-d), we show both the initial and final configuration of an experiment with 100\% river sand particles. The initial height was also 50 mm, and the final run-out distance was approximately 68 mm. Figure \ref{fig:exp_setup}(e-f) shows an example of mixing glass and sand. The mass ratio between two types of particles was $1:1$. Given their similar density, the volume ratio was also approximately $1:1$. In Figure \ref{fig:exp_setup}(e-f), the initial height was 50 mm, and the final run-out distance was approximately 75 mm.

\begin{figure}
  \centerline{\includegraphics[scale=0.4]{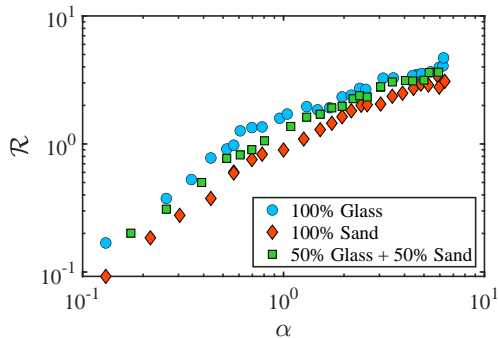}}
  \caption{Experimental results of the relationship between the initial aspect ratio, $\alpha$, and the relative run-out distance, $\mathcal{R}$.}
\label{fig:exp_results}
\end{figure}

As shown in Figure \ref{fig:exp_setup}, changing the mixing ratio affects the final run-out distance. Thus, replacing part of sand particles with glass particle improves the mobility of the granular mixture. We further tested columns with different initial height and plot the relationship between $\mathcal{R}$ and $\alpha$ in Figure \ref{fig:exp_results}. As we increase the percentage of the glass particles in the mixture, the relative run-out distance becomes larger. The experimental results agreed with our expectation, and implied that we should develop a method to quantify such mixing effect.

\begin{figure}
  \centerline{\includegraphics[scale=0.4]{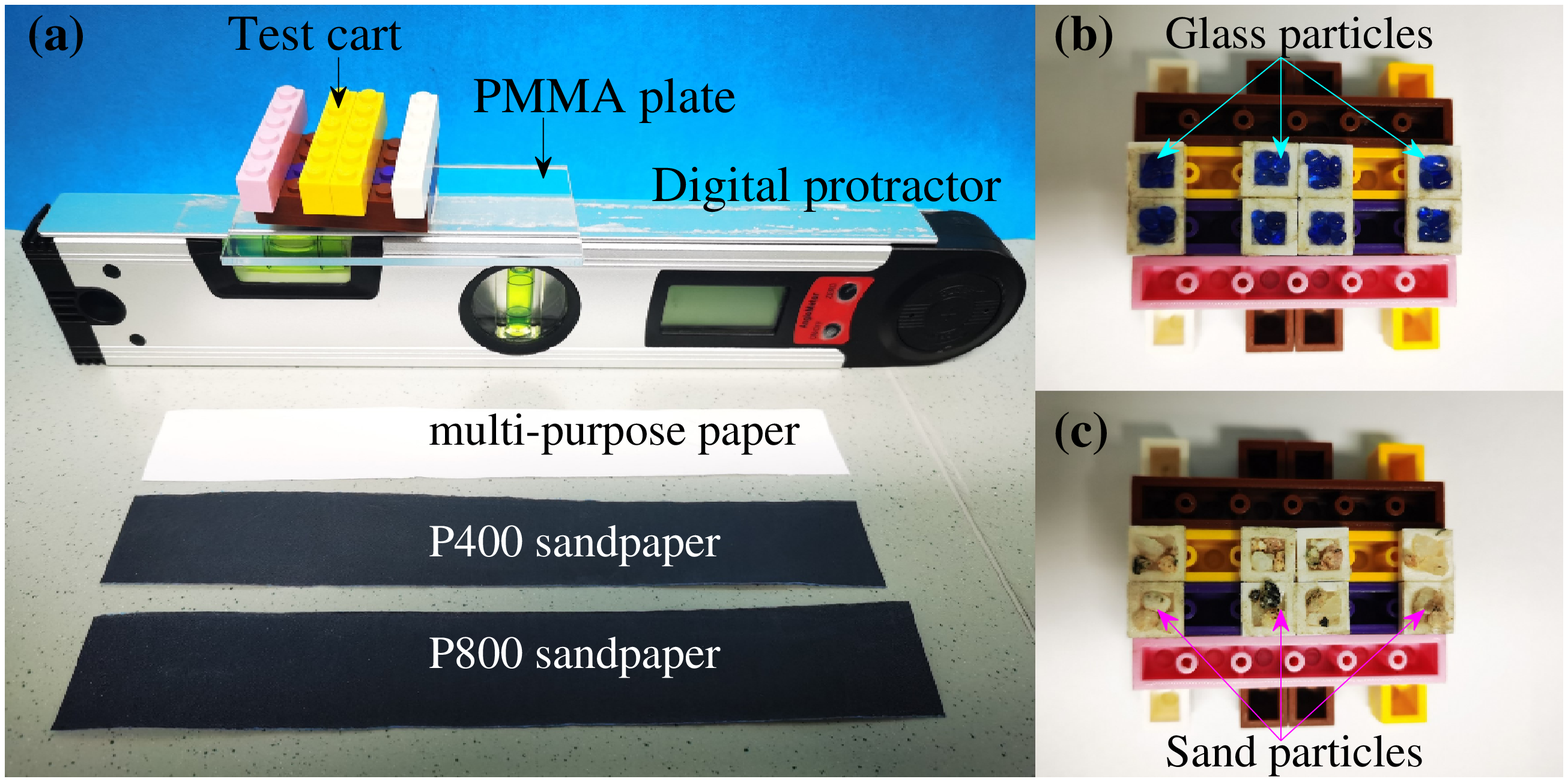}}
  \caption{Experimental setup for measuring the frictional coefficient between sand or glass particles and different basal materials.}
\label{fig:fricSetup}
\end{figure}

\begin{table}
  \begin{center}
\def~{\hphantom{0}}
  \begin{tabular}{lcc}
      Bottom material & Glass particles & River Sand \\[3pt]
       PMMA plate & 0.075$\pm$0.01 & 0.097$\pm$0.006\\
       80 g/cm$^2$ copypaper & 0.058$\pm$0.003 & 0.177$\pm$0.011\\
       P400 sandpaper & 0.194$\pm$0.007 & 0.246$\pm$0.012\\
       P800 sandpaper & 0.189$\pm$0.004 & 0.224$\pm$0.018\\
  \end{tabular}
  \caption{Frictional coefficients (the average $\pm$ standard deviation) between experimental particles and different types of bottom plates.}
  \label{tab:fric}
  \end{center}
\end{table}

Our analyses of the experimental results are based on the assumption that sand particles are generally rougher than glass particles; hence, have larger frictional coefficient. To verify this assumption, we need more concrete experimental results rather than only using our physical intuition. It is difficult to directly test the frictional coefficient between particles \citep{foerster1994,lorenz1997}. Thus, we choose to test the frictional coefficient between particles and different types of bottom plates. We build up the friction test platform using only LEGO$\circledR$ blocks, a plastic (PMMA) plate, a digital protractor, 80 g/cm$^2$ multi-purpose copy paper, and two types of sandpapers, as shown in Figure \ref{fig:fricSetup}(a). We use LEGO$\circledR$ blocks to form a testing cart, where we could place testing particles beneath the cart to form several non-rolling particle ``feet" [Figure \ref{fig:fricSetup}(b) and (c)]. We place the testing cart with granular ``feet" onto certain basal plates of different materials (we make sure the basal materials are glued firmly onto the protractor) and lift one end of the plate until the testing cart starts to move. Since the testing plate was placed on a digital protractor, we could easily measure the inclined angle, $\theta_s$, when the cart starts to move. Thus, the tested frictional coefficient between targeting granular materials and the basal materials is calculated as $\mu_{pb} = \textrm{tan}(\theta_s)$.  We present friction test results of the average frictional coefficient and the standard deviation in Table \ref{tab:fric}, where we measure frictional coefficients using a smooth PMMA plate, 80 g/cm$^2$ multi-purpose papers, P400 sandpapers, and P800 sandpapers (smoother than P400 sandpapers). Generally, the average frictional coefficients of glass particles on different base materials are smaller than those of sand particles. 

We have to note that experimental results cannot confirm the influence of mixing frictions and quantify its influence. On one hand, we merely show that sand particles are generally rougher than glass particles. On the other hand, smaller frictional coefficients lead to denser initial packing. For the same initial aspect ratio, glass particle packing often has a larger number of particles due to its denser packing condition, which may result in larger run-out distance in the end. After showing the evidence of the influence of mixing particles with different frictional coefficient, we naturally move to simulation tools so that we can control the parameters much more easily.

\section{Simulations}
\label{sec:simu}
\subsection{Governing equations}
\label{DEM}
In this study, we performed simulations with the discrete element method (DEM) \citep{galindo2010molecular} to test the collapse of granular columns with different frictional coefficients, which allowed us to easily and specifically control certain parameter and to extract particle-scale data from the system. We use Voronoi-based sphero-polyhedral particles in our simulations. The sphero-polyhedra method was initially introduced by \citet{pournin2005generalization} for the simulation of complex-shaped DEM particles. A sphero-polyhedron is a polyhedron that has been eroded and then dilated by a spherical element. The result is a polyhedron of similar dimensions but with rounded corners.


The advantage of the sphero-polyhedral technique is its easy and efficient definition of contact relationships among particles. When we calculate the contact force between adjacent particles, we can directly use the contact between their dilating spheres. Then, the contact calculation of complex-shaped particles is transformed into the contact between spheres. For example, we consider the contact between two generic particles named $P_1$ and $P_2$. $P_1$ has geometric features, such as a set of vertices, $\{V_{1}^i\}$, edges $\{E_{1}^j\}$, and faces $\{F_{1}^k\}$. $P_2$ also has geometric features, such as a set of vertices, $\{V_{2}^i\}$, edges $\{E_{2}^j\}$, and faces $\{F_{2}^k\}$. Thus, a particle is defined as a polyhedron, i.e. a set of vertices, edges and faces, where each one of these geometrical feature is dilated by a sphere. For simplicity, we denote the set of all the geometric features of $P_1$ and $P_2$ as $\{G_{1}^i\}$ and $\{G_{2}^j\}$. Then, we can calculate the distances between $\{G_{1}^i\}$ and $\{G_{2}^j\}$ as
\begin{equation}\label{eq:dist}
 \textrm{dist}(G_1^i,G_2^j) = \min\left (\textrm{dist}(\vec{X}_1^i,\vec{X}_1^j)\right ),
\end{equation}
where $\vec{X}_1^i$ is a 3D vector of points that belongs to the set $G_1^i$ and $\vec{X}_2^j$ is a 3D vector of points that belongs to the set $G_2^j$. This means that the distance for two geometric features is the minimum Euclidean distance assigned to two points belonging to them.

Since both particles are dilated by their sphero-radii $R_1$ and $R_2$, a contact is confirmed when the distance between the two geometric features is less than the addition of the corresponding radii used in the sweeping stage, i.e.,
\begin{equation}\label{eq:contact}
 \textrm{dist}(G_1^i,G_2^j) < R_1+R_2,
\end{equation}
and the corresponding contact overlap $\delta_n$ can be calculated accordingly. Thus, the advantage of the sphero-polyhedra technique becomes evident since this definition is similar to the one for the contact law of two spheres \citep{belheine2009numerical}. For each confirmed contact, we implement a Hookean contact model with energy dissipation to calculate the interactions between particles. At each time step, the overlap between adjacent particles, $\delta_n$, is checked and the normal contact force can be calculated using
\begin{equation}
        \vec{F}_n = -K_n\delta_n\hat{n} - m_e\gamma_{n}\vec{v}_n,
\end{equation}
where $K_n$ is the normal stiffness characterizing the deformation of the material, $\hat{n}$ is defined as the normal unit vector at the plane of contact, $\vec{v}_n$ is the relative normal velocity between particles, $m_e = 0.5(1/m_1 + 1/m_2)$ is the reduced mass of the contacting particle pair, $m_1$ and $m_2$ are masses of contacting particles, respectively, and $\gamma_n$ is the normal energy dissipation constant, which depends on the coefficient of restitution $e$ as \citep{alonso2013experimental,galindo2018micromechanics},
\begin{equation} \label{eq:restitution}
        e = \textrm{exp}\left(-\frac{\gamma_n}{2} \frac{\pi}{\sqrt{\frac{K_n}{m_e} - (\frac{\gamma_n}{2})^2}}\right)\ .
\end{equation}


The tangential contact forces between contacting particles were calculated by keeping track of the tangential relative displacement $\vec{\xi} = \int \vec{v}_tdt$. Thus, the tangential contact forces follow
\begin{equation} \label{eq:contactforce}
        \vec{F}_t = -\textrm{min}\left(|K_t\vec{\xi}|,\ \mu_p |\vec{F}_n|\right)\hat{t},
\end{equation}
where $K_t$ is the tangential stiffness, $\hat{t}$ is the tangential vector in the contact plane and parallel to the tangential relative velocity, $\vec{v}_t$, and $\mu_p$ is the frictional coefficient between contacting particles and can be replaced by the frictional coefficient between the particles and the bottom boundary, $\mu_w$, while calculating the particle-boundary interactions. In this study, since we use Voronoi-based particles, no rolling resistance is needed. The motion of particles is then calculated by step-wise resolution of Newton's second law with the normal and contact forces mentioned before, so that
\begin{subeqnarray}\label{eq:governEqn}
    m_p\frac{\textrm{d}^2\textbf{X}_p}{\rm{d}t^2} &=& \sum_c^{N_c}\left( \textbf{F}_n^{pc} + \textbf{F}_t^{pc} \right)\ , \\[3pt]
    \frac{\textrm{d}}{\textrm{d}t}\left(\textbf{I}_p \boldsymbol\omega_p \right) &=& \textbf{T}_t,
\end{subeqnarray}
where $\textbf{X}_p$ is the position vector of a particle, $m_p$ is the mass of a particle, $N_c$ is the number of contacts, $\textbf{F}_n^{pc}$ and $\textbf{F}_n^{pc}$ are normal and tangential contact force vectors acting from the contact point to the particle, $\textbf{I}_p$ is the tensor of the moment of inertial of the particle, $\boldsymbol\omega_p$ is the angular velocity vector of the particle, and $\textbf{T}_t$ is the total torque subjected to the particle. 

In the DEM simulation, we solve the governing equations of this classical interacting N-body system using the velocity-Verlet method \citep{Scherer2017}.The same neighbor detection and force calculation algorithms have already been discussed and validated in previous studies \citep{galindo2010molecular,man2021deposition}, and the presented DEM formulation has been validated before with experimental data \citep{belheine2009numerical,cabrejos-hurtado2016} and is included in the MechSys open source multi-physics simulation library \citep{galindo2013coupled}. 

\subsection{Simulations setup}

\begin{figure}
  \centerline{\includegraphics[scale=0.45]{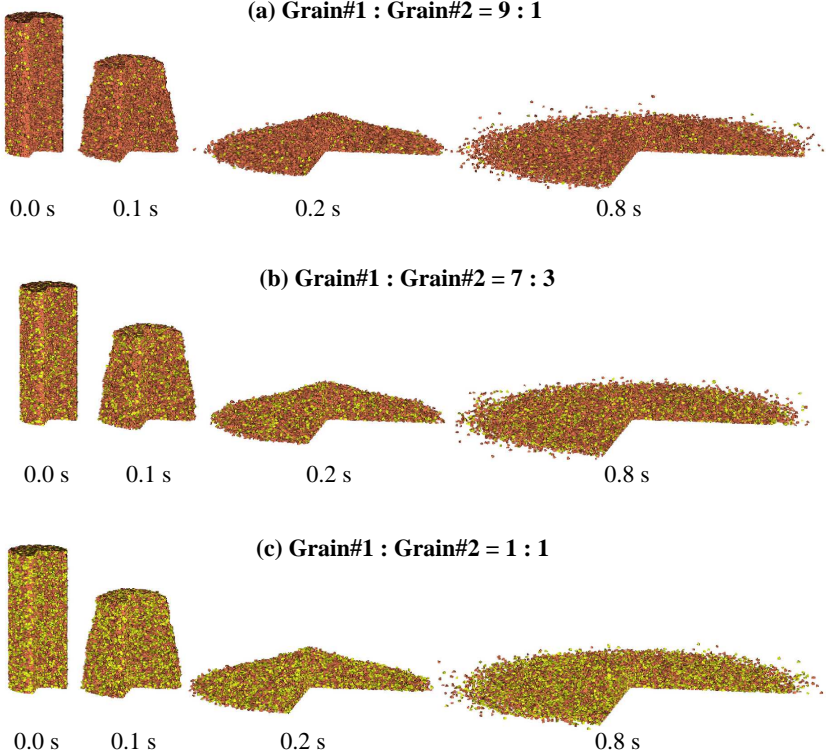}}
  \caption{Simulation setup and collapse behaviors of systems with different mixing ratios. Red particles represent Grain\#1 and yellow particles represent Grain\#2. We cut one quarter of the granular assembly to show the inside of the system.}
\label{fig:simuSetup}
\end{figure}

We performed simulations of the granular column collapses with Voronoi-based sphero-polyhedra \citep{galindo2010molecular}. We note that the shape of particles could significantly influence of the deposition morphology. In this study, we focus on using Voronoi-based particles so that the particles in the simulation are similar to sand particles. The detailed influence of particle shapes on the granular column collapses will be further explored in the future. In a simulation, we first generate Voronoi-based particle packing in a designed cylindrical domain of height $H_i$ and radius $R_i = 2.5$ cm (Figure \ref{fig:simuSetup}). The number of particles within one unit length (1.0 cm) is 5, so the average particle size is $\approx 2$ mm. Particles were packed within a column of radius $R_i$ equal to 2.5 cm and varying heights $H_i$ leading to cases of different initial aspect ratio. Then, 20\% of the sphero-polyhedron particles were removed to form a packing with a solid fraction of $\phi_s =$ 0.8. $H_i$ varies from 1 cm to 40 cm. In the simulations of Voronoi-based particles, the number of particles varied from approximately 1900 to approximately 68500. The initial state of the granular column resembles a fissured rock with initial solid fraction $\phi_s = 0.8$. Then, we removed the cylindrical tube in the simulation and let grains flow downward freely with the gravitational acceleration $g = 981$ cm/s$^2$ (Figure \ref{fig:simuSetup}). Figure \ref{fig:simuSetup} shows the behavior of a granular column from the initial state to the final deposition state. Each row of Figure \ref{fig:simuSetup} represents a granular column with distinct mixing ratio of Grain\#1 and Grain\#2 (Grain\#1 and Grain\#2 only differ in their frictional properties). In the end, a cone-like pile of granular material with packing height, $H_{\infty}$, and average packing radius, $R_{\infty}$, will form . 

We implemented the Hookean contact model (elaborated in Sect. \ref{DEM}) with energy dissipation and restitution coefficient $e = 0.1$ to calculate the interactions between particles as we have described in the above section. A relatively low value of $e$ was chosen to represent the rough surface of particles in real conditions \citep{li2020surface}. We introduce two species of Voronoi-based particles in a system, where frictional properties of Grain\#1 and Grain\#2 are set separately. The frictional coefficient of the contact between Grain\#1 and Grain\#1, $\mu_{11}$ varies from 0.1 to 0.8.  Similarly, we vary the frictional coefficient of the contact between Grain\#2 and Grain\#2, $\mu_{22}$, from 0.1 to 0.8. The frictional coefficient between Grain\#1 and Grain\#2 is then calculated as
\begin{equation} \label{eq:mu12}
    \mu_{12} = \frac{2\mu_{11}\mu_{22}}{\mu_{11} + \mu_{22}}.
\end{equation}

Simulations were conducted with varied initial aspect ratios, $\alpha$, between 0.4 and 16, varied mixing ratio where the percentage of Grain\#2 varies from 10\% to 50\%, and a constant particle/boundary frictional coefficient, $\mu_w = 0.4$, which is the same to both Grain\#1 and Grain\#2. Based on these simulations we obtained the run-out behavior and deposition morphology for different conditions.

\section{Results and discussions}
\label{sec:res_discuss}

\subsection{Flow behavior}

\begin{figure}
  \centerline{\includegraphics[scale=0.35]{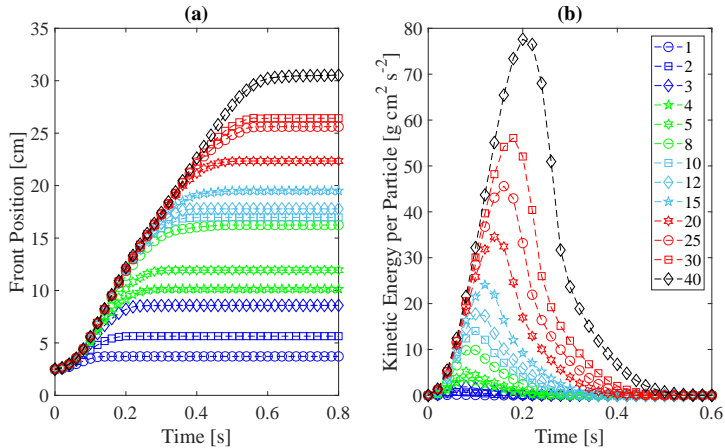}}
  \caption{(a) Relationship between the front position and time during the collapse of granular systems with different initial height (shown in the legend in centimeter). (b) Shows the relationship between the average particle kinetic energy and time. In this figure, Grain\#2 makes up of 10\% of the total number of particles. $\mu_{11} = 0.1$, $\mu_{22} = 0.4$, and $\mu_w = 0.4$.}
\label{fig:flowBehav10}
\end{figure}

\begin{figure}
  \centerline{\includegraphics[scale=0.35]{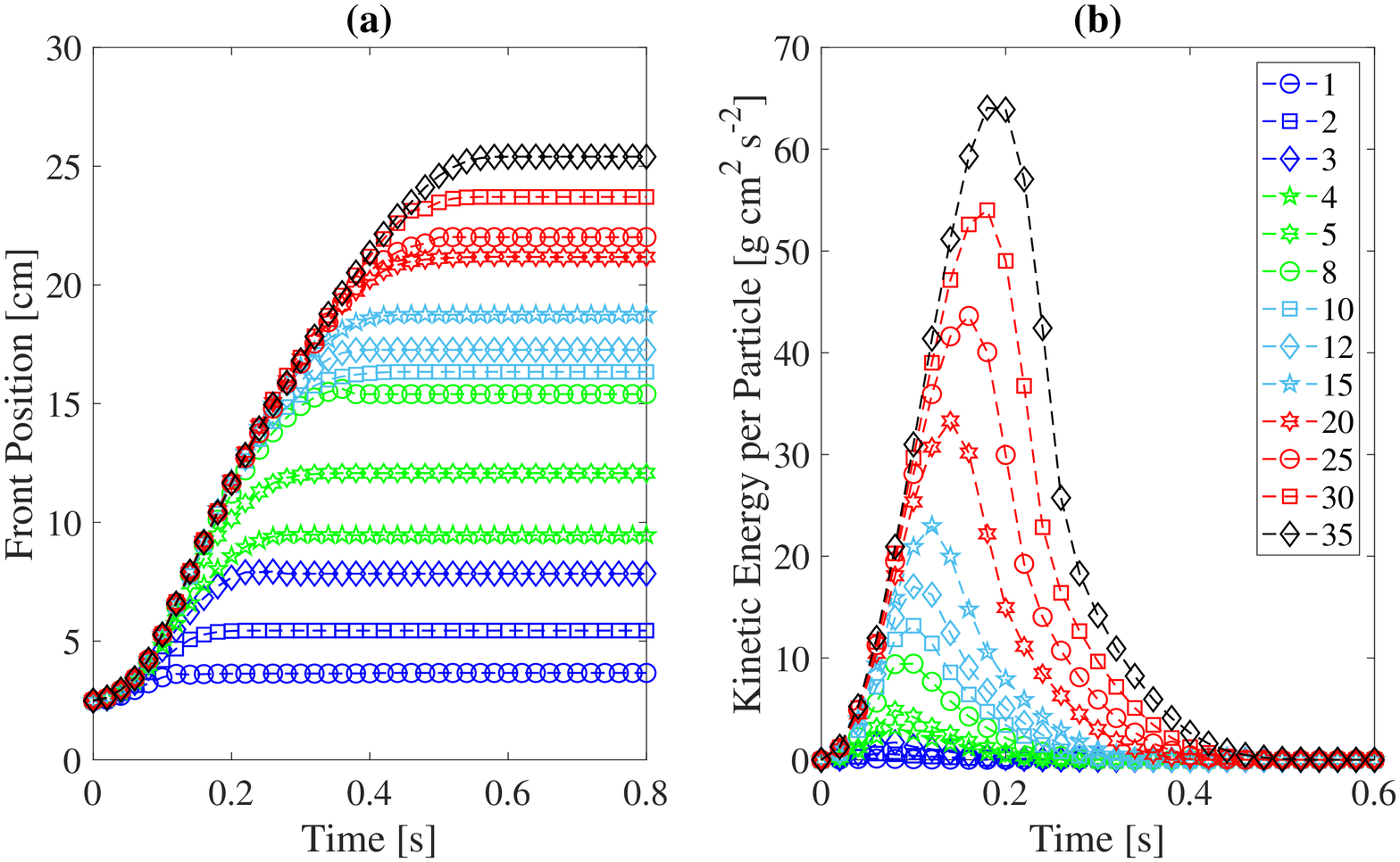}}
  \caption{(a) Relationship between the front position and time during the collapse of granular systems with different initial height (shown in the legend). (b) Shows the relationship between the average particle kinetic energy and time. In this figure, Grain\#2 makes up of 30\% of the total number of particles. $\mu_{11} = 0.1$, $\mu_{22} = 0.4$, and $\mu_w = 0.4$.}
\label{fig:flowBehav30}
\end{figure}

\begin{figure}
  \centerline{\includegraphics[scale=0.35]{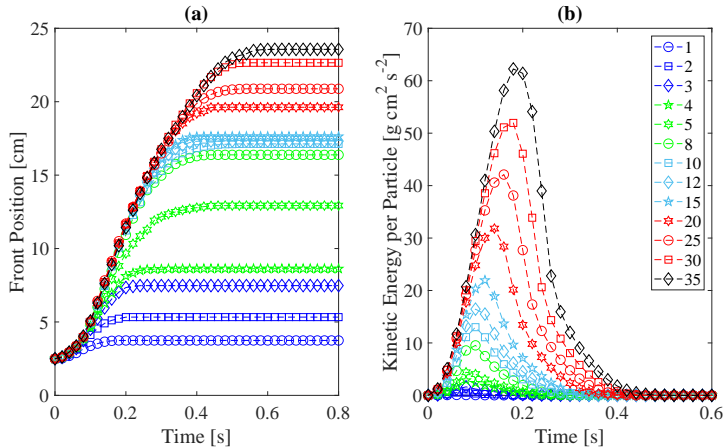}}
  \caption{(a) Relationship between the front position and time during the collapse of granular systems with different initial height (shown in the legend). (b) Shows the relationship between the average particle kinetic energy and time. In this figure, Grain\#2 makes up of 50\% of the total number of particles. $\mu_{11} = 0.1$, $\mu_{22} = 0.4$, and $\mu_w = 0.4$.}
\label{fig:flowBehav50}
\end{figure}

Generally, based on the propagation velocity of the front, a granular column collapse can be divided into three stages: (1) the acceleration stage, (2) the steady-propagating stage, and (3) the deceleration stage. In Figures \ref{fig:flowBehav10} - \ref{fig:flowBehav50}, we measure the front position and the average kinetic energy for three sets of simulations and plot them against the collapse time. These three set of simulations have different mixing ratios, but the same frictional coefficients ($\mu_{11} = 0.1$, $\mu_{22} = 0.4$). The resulting front positions behave similarly among three set of simulations. As we increase the initial height of the granular column, the time when the granular flow stops varies from case to case. Columns with larger initial height can travel for longer time, since they need more time to dissipate the stored potential energy. We hypothesize that there may exist a relationship between the effective aspect ratio and the terminal time, $t_f$, when the system stop flowing.

The relationship between the average kinetic energy and the time shows that systems with different initial height reach their maximum kinetic energy at different time, $t_{\rm{max}}$. For instance, a granular column with $H_i = 1.0$ cm often reaches the maximum kinetic energy at $t_{\rm{max}}\approx 0.06$s, but a granular column with $H_i = 35$ cm reaches its maximum kinetic energy at $t_{\rm{max}}\approx 0.2$s. Similarly, we may also obtain a relationship between $t_{\rm{max}}$ and $\alpha_{\rm{eff}}$.

Changing the mixing ratio also influences the collapse behavior, even though Figures \ref{fig:flowBehav10}-\ref{fig:flowBehav50} do not differ much from each other. Take simulations with $H_i = 30$ cm for example. When Grain\#1 : Grain\#2 = 9 : 1 as shown in Figure \ref{fig:flowBehav10}, $t_{\rm{max}}\approx 0.18$ and $t_f\approx 0.56$. When Grain\#1 : Grain\#2 = 7 : 3 as shown in Figure \ref{fig:flowBehav30}, $t_{\rm{max}}\approx 0.18$ but $t_f\approx 0.54$. When Grain\#1 : Grain\#2 = 1 : 1 as shown in Figure \ref{fig:flowBehav50}, $t_{\rm{max}}\approx 0.18$ (slightly $< 0.18$) but $t_f\approx 0.52$. Since we extract data every 0.02 seconds, $t_{\rm{max}}$ may not be accurate enough, but $t_f$ certainly shows the trend that having more rough particles in a system decreases the terminal time, $t_f$. Changing mixing ratios also affects the maximum kinetic energy a granular system can reach. Taking simulation results of systems with $H_i = 30$ cm for example, the maximum kinetic energy per particle can reach $\approx 58$ g cm$^2$s$^{-2}$ for a granular column with mixing ratio equal to $9:1$, while the maximum kenetic energy per particle can only reach $\approx 52$ g cm$^2$s$^{-2}$ when mixing ratio is $9:1$, as we keep other parameters the same. The detailed analyses of $t_{\rm{max}}$ and $t_{\rm{f}}$ will be presented in Sect. \ref{sec:kinematic} and \ref{sec:further}.

\subsection{Run-out distances}
\label{sec:runout}
\begin{figure}
  \centerline{\includegraphics[scale=0.32]{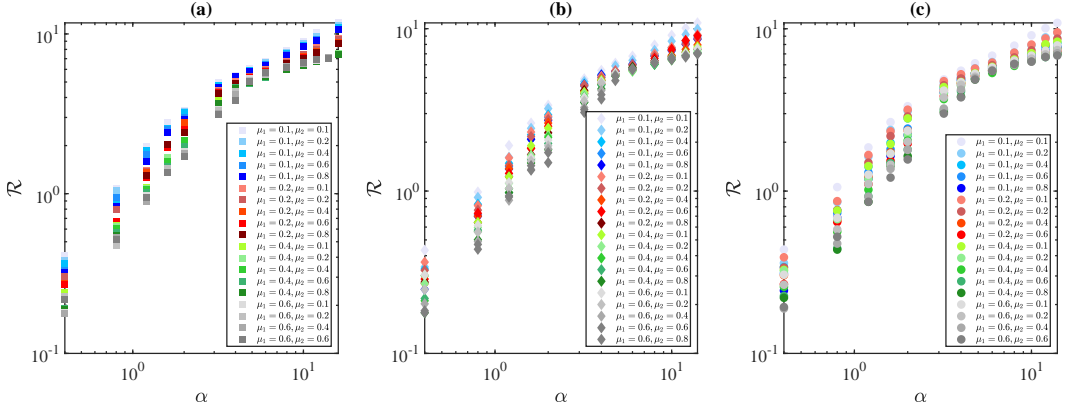}}
  \caption{Simulation results of the relationship between the relative run-out distance, $\mathcal{R}$, and the initial aspect ratio, $\alpha$.}
\label{fig:R_alpha}
\end{figure}

Figure \ref{fig:R_alpha} shows the relationship between the relative run-out distance, $\mathcal{R}$, and the initial aspect ratio, $\alpha$, of systems with different frictional coefficients. Figure \ref{fig:R_alpha}(a), (b), and (c) have different mixing ratios, but their behaviors look similar. For granular columns with the same frictional property, varying the initial aspect ratio results in two regimes of granular column collapses with a critical aspect ratio, $\alpha_c$, that, when $\alpha<\alpha_c$, $\mathcal{R}$ scales approximately with $\alpha$, and when $\alpha>\alpha_c$, $\mathcal{R}$ scales approximately with $\alpha^{0.5}$, as first determined for a mono-particle system by \citet{lube2004axisymmetric}. Also, similar to the work of \citet{man2021deposition}, decreasing the frictional coefficient increases the relative run-out distance. As shown in Figure \ref{fig:R_alpha}, the light blue markers, which represent granular systems with small frictional coefficients, always locate above other markers. 

\begin{figure}
  \centerline{\includegraphics[scale=0.35]{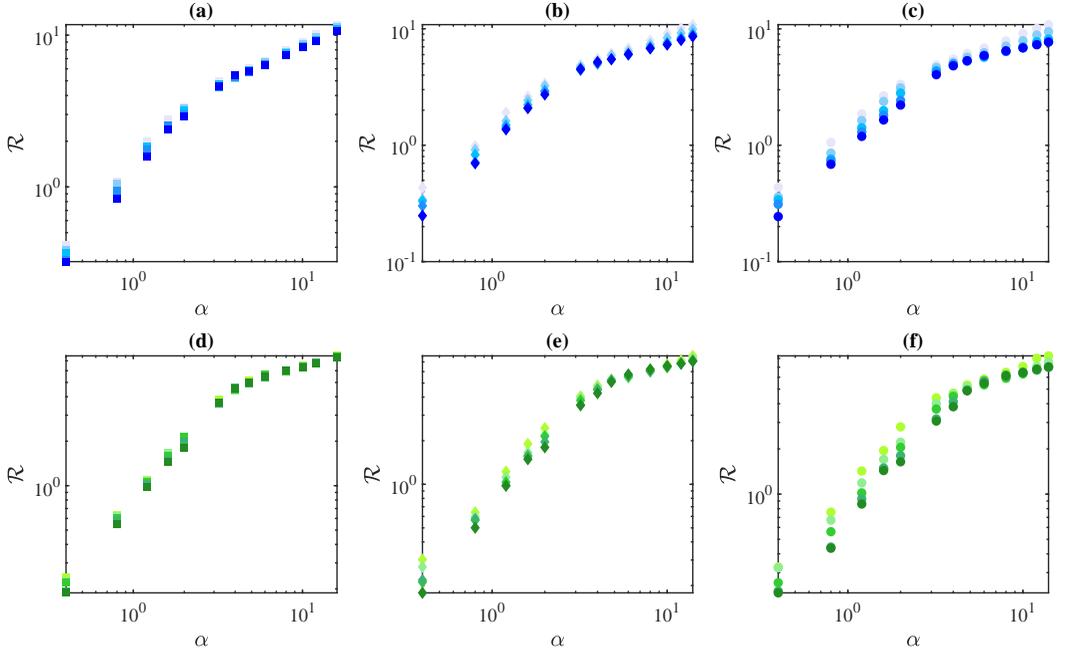}}
  \caption{relationship between the relative run-out distance, $\mathcal{R}$, and the initial aspect ratio, $\alpha$, of selected sets of simulations to gain more detailed information, where Figures (a - c) have the same $\mu_{11} = 0.1$ but different mixing ratios of $9:1$, $7:3$, and $1:1$, and Figures (d - f) have the same $\mu_{11} = 0.4$ but different mixing ratios of $9:1$, $7:3$, and $1:1$. In each of the sub-figure, we vary the initial height from 1 cm to 40 cm and $\mu_{22}$ from 0.1 to 0.8. Markers in these figures are the same as those in Figure \ref{fig:R_alpha}.}
\label{fig:R_alpha2}
\end{figure}

The mixing ratio can also influence the behavior of run-out distances, since changing mixing ratio inevitably affects the bulk frictional property of the system. We extract six sets of simulation results to show the influence of mixing ratios and plot them in Figure \ref{fig:R_alpha2}. Figures \ref{fig:R_alpha2}(a - c) show simulations with $\mu_{11} = 0.1$ and $\mu_{22} = 0.1, 0.2, 0.4, 0.6, 0.8$. Figures \ref{fig:R_alpha2}(d - f) plot the relationship between $\mathcal{R}$ and $\alpha$ for systems with $\mu_{11} = 0.4$ and $\mu_{22} = 0.1, 0.2, 0.4, 0.6, 0.8$. The ratios between Grain\#1 and Grain\#2 are 9:1 [Figure \ref{fig:R_alpha2}(a,d)], 7:3 [Figure \ref{fig:R_alpha2}(b,e)], and 1:1 [Figure \ref{fig:R_alpha2}(c,f)], respectively. We can see in Figures \ref{fig:R_alpha2}(a) and (d) that, when the mixing ratio is $9:1$, changing the frictional coefficient of Grain\#2 without changing the friction of Grain\#1 brings little impact on the $\mathcal{R}(\alpha)$ curve since Grain\#1 accounts for 90\% of all the particles and the Grain\#1-Grain\#1 interaction should be dominant during the collapse. However, as we increase the percentage of Grain\#2, the constant frictional coefficient among Grain\#1 starts to lose its dominance in the collapse. As shown in Figures \ref{fig:R_alpha2}, when the percentage of Grain\#2 is equal to that of Grain\#1, changing the frictional coefficient among Grain\#2, $\mu_{22}$, while keeping $\mu_{11}$ constant, has more influence, resulting in a larger spread width in the $\mathcal{R}(\alpha)$ plot.

\begin{figure}
  \begin{subfigure}{1.0\textwidth}
    \centering
    \includegraphics[scale=0.36]{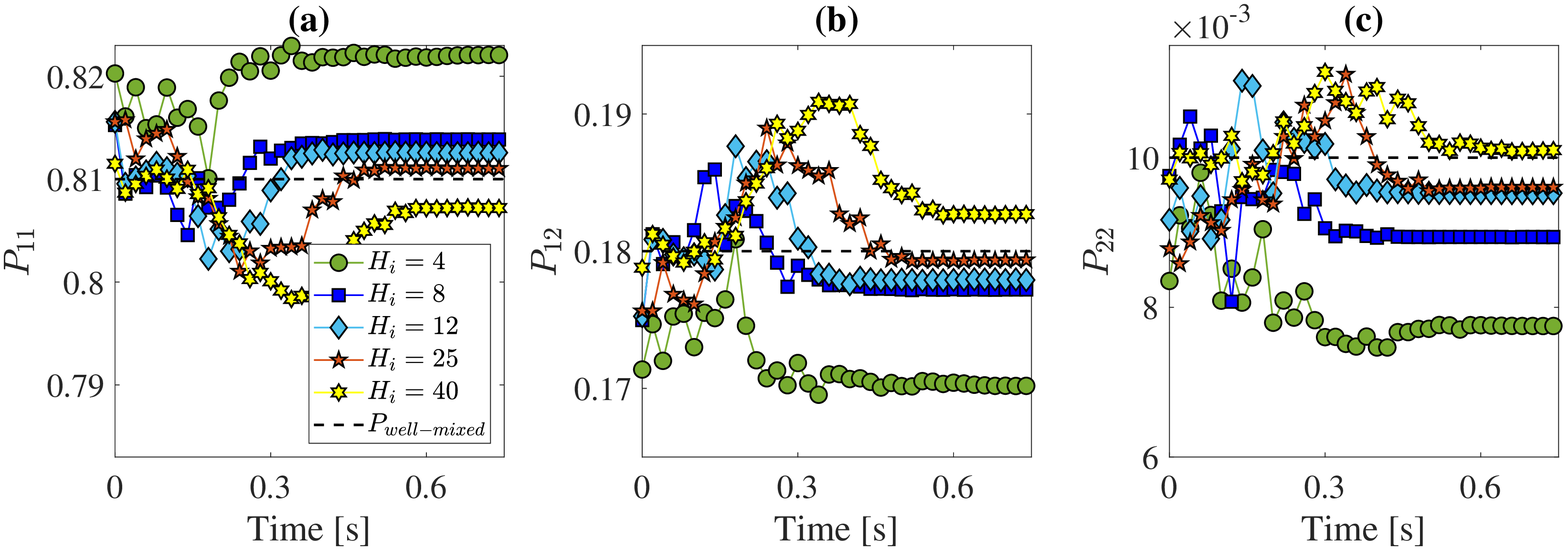}
  \end{subfigure}
  \newline
  \begin{subfigure}{1.0\textwidth}
    \centering
    \includegraphics[scale=0.36]{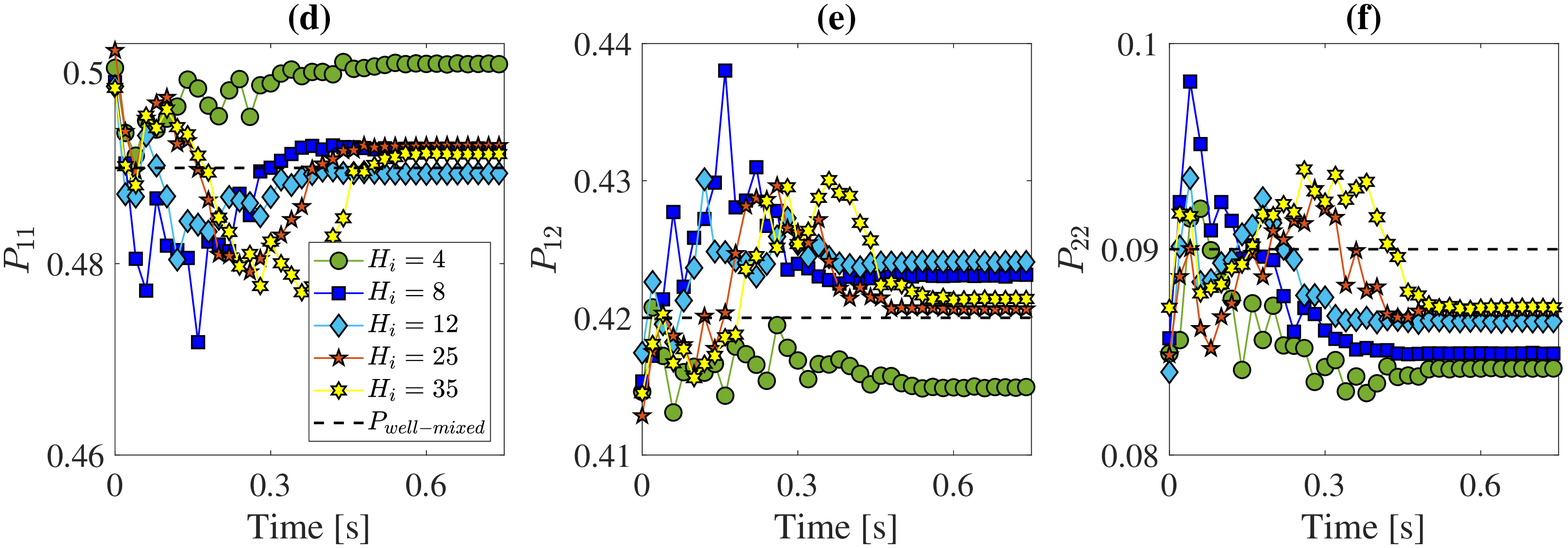}
  \end{subfigure}
  \newline
  \begin{subfigure}{1.0\textwidth}
    \centering
    \includegraphics[scale=0.36]{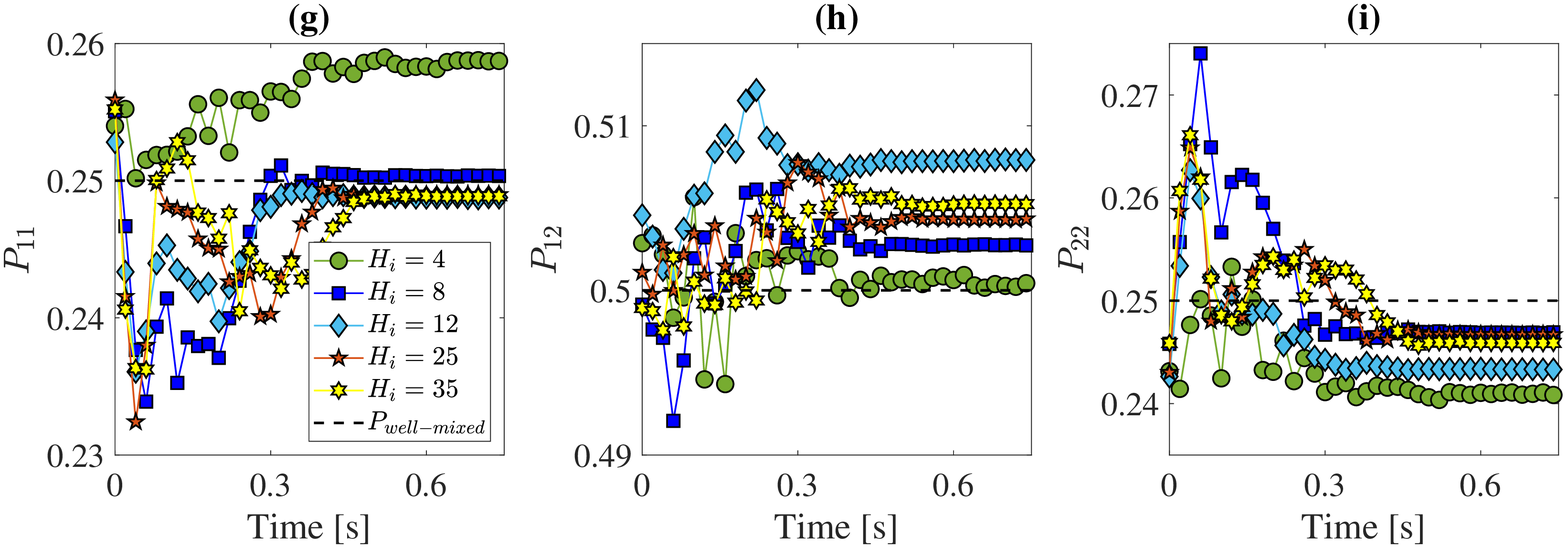}
  \end{subfigure}
  \caption{Evolution of contact occurrence probability of Grain\#1-Grain\#1 contact ($P_{11}$), Grain\#1-Grain\#2 contact ($P_{12}$), and Grain\#2-Grain\#2 contact ($P_{22}$) with respect to collapsing time. Figures (a - c) show probability of systems with $\mu_{11} = 0.1$, $\mu_{22} = 0.6$, and Grain\#1 : Grain\#2 = 9 : 1, and they share the same legend as presented in Figure (a). Figures (d - f) show probability of systems with $\mu_{11} = 0.1$, $\mu_{22} = 0.6$, and Grain\#1 : Grain\#2 = 7 : 3, and they share the same legend as presented in Figure (d). Figures (g - i) show probability of systems with $\mu_{11} = 0.1$, $\mu_{22} = 0.6$, and Grain\#1 : Grain\#2 = 1 : 1, and they share the same legend as presented in Figure (g).}
\label{fig:contactPerc}
\end{figure}

Figures \ref{fig:R_alpha} and \ref{fig:R_alpha2} show the influence of friction and the influence of the mixing ratio. We hypothesize that the influence of the mixing ratio is related to the contact probability of three existing contact in the system: (1) Grain\#1-Grain\#1 contact; (2) Grain\#1-Grain\#1 contact; (3) Grain\#1-Grain\#1 contact. When a system is well-mixed and has infinite number of particles of two different species, where the percentage of Grain\#1 is $P_1$ and the percentage of Grain\#2 is $P_2 = 1-P_1$, the contact probability of each contact type can be well defined that
\begin{subeqnarray} \label{eq:contactPerc}
    P_{11} &=& P_1^2,\ \ P_{22} = P_2^2 = (1 - P_1)^2,\\[3pt]
    P_{12} &=& 2P_1P_2 = 2P_1(1-P_1),
\end{subeqnarray}
where $P_{11}$ is the percentage of Grain\#1-Grain\#1 contact among all the contact pairs, $P_{22}$ is the percentage of Grain\#2-Grain\#2 contact, and $P_{12}$ is the percentage of Grain\#1-Grain\#2 contact.

During granular column collapses, the system is subjected to shearing deformation. In our previous work \citep{man2023friction}, we concluded that inter-particle frictional coefficient influences the rheological behavior of the sheared granular assembly. Thus, we are uncertain about the existence of segregation effect during the column collapse, which may change of the percentage of each contact type. In Figure \ref{fig:contactPerc}, we plot the percentage of each contact type for systems with different initial heights and different mixing ratios. Figures \ref{fig:contactPerc}(a - c) show the contact percentage for systems with $\mu_{11} = 0.1$, $\mu_{22} = 0.6$, and Grain\#1 : Grain\#2 = 9 : 1. Since $P_1 = 0.9$ and $P_2 = 0.1$, we expect that $P_{11} = 0.81$, $P_{12} = 0.18$, and $P_{22} = 0.01$. Even though $P_{11}$, $P_{12}$, and $P_{22}$ changes with different initial height and different measuring time, they do not deviate much from the theoretical value (shown in Figure \ref{fig:contactPerc} as black dashed lines, which show that, both at the initial state and during the collapse, the system remains well-mixed and no obvious segregation happens during the collapse. Similar behavior can be observed for systems with Grain\#1 : Grain\#2 = 7 : 3 [Figures \ref{fig:contactPerc}(d - f)] and Grain\#1 : Grain\#2 = 1 : 1 [Figures \ref{fig:contactPerc}(g - i)].

\begin{figure}
  \centerline{\includegraphics[scale=0.4]{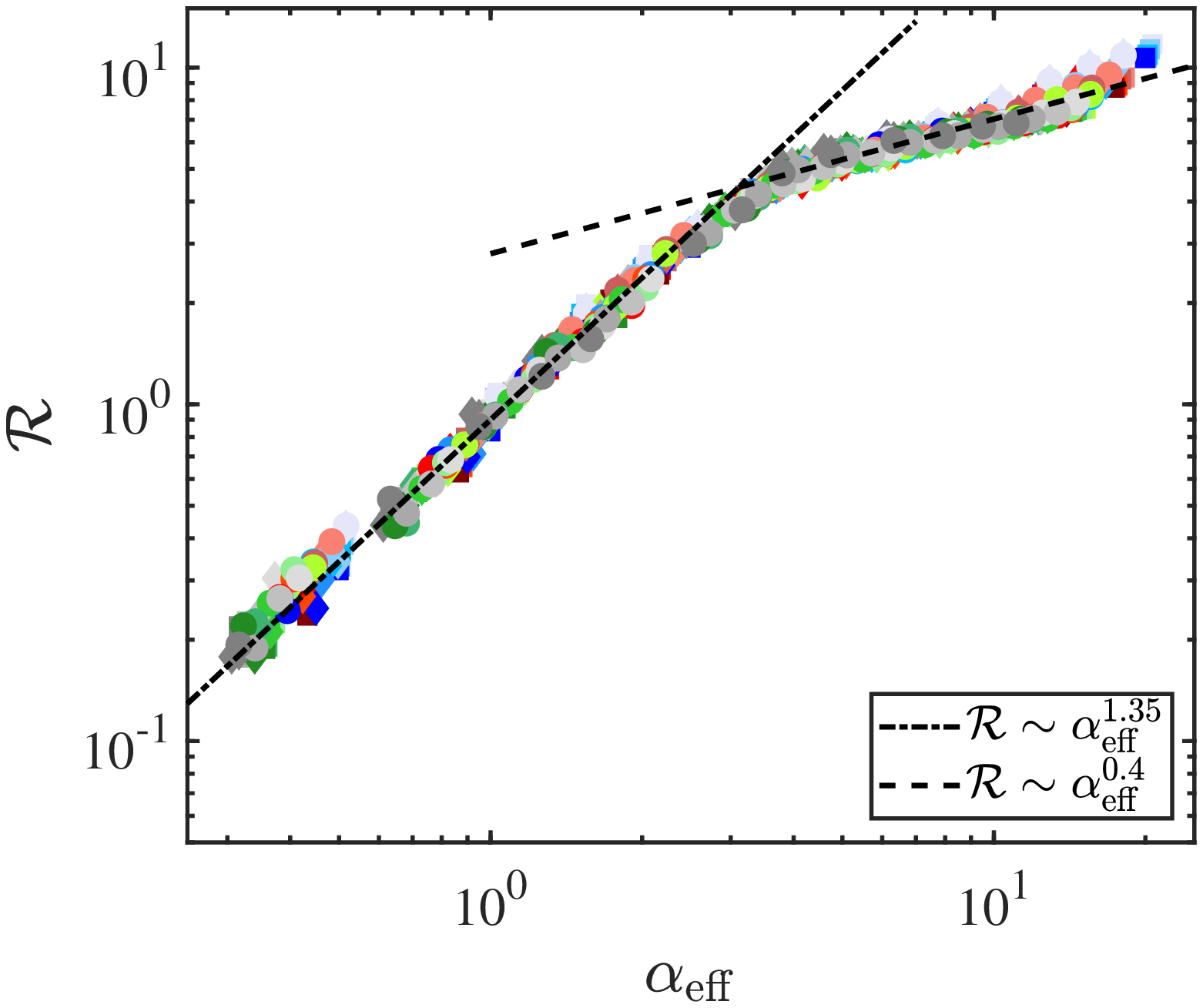}}
  \caption{relationship between the relative run-out distance, $\mathcal{R}$, and the effective aspect ratio, $\alpha_{\rm{eff}}$. Markers in this figures are the same as those in Figure \ref{fig:R_alpha}.}
\label{fig:R_alphaeff}
\end{figure}

In Eqn. \ref{eq-alpha-eff} and \citet{man2021deposition}, we state that the effective aspect ratio, obtained from dimensional analysis and including the influence of frictional coefficient, helps unify the $\mathcal{R}(\alpha_{\rm{eff}})$ relationship. In Eqn. \ref{eq-alpha-eff}, the influence of friction can be divided into two parts: one is the particle-boundary friction, $\mu_w$, and the other one is the inter-particle friction, $\mu_p$. In this work, we argue that $\mu_p$ can be further decomposed into three different contact types, since there exists two different species of particles. The frictional coefficient between Grain\#1 and Grain\#1 is $\mu_{11}$, and its occurrence probability is $P_{11}$. The frictional coefficient between Grain\#2 and Grain\#2 is $\mu_{22}$, and its occurrence probability is $P_{22}$. Similarly, the frictional coefficient between Grain\#1 and Grain\#2 is $\mu_{12} = 2\mu_{11}\mu_{22}/(\mu_{11} + \mu_{22})$, and its occurrence probability is $P_{12}$. A simple mixture theory enable us to write the general inter-particle frictional coefficient, $\mu_p$, and the effective aspect ratio, $\alpha_{\rm{eff}}$, as
\begin{subeqnarray}
    \mu_p &=& \mu_{11}P_{11} + \mu_{22}P_{22} \mu_{12}P_{12}, \\[3pt]
    \alpha_{\rm{eff}} &=& \alpha\sqrt{1/\left[\mu_w + \beta(\mu_{11}P_{11} + \mu_{22}P_{22} + \mu_{12}P_{12}) \right]}
\end{subeqnarray}
where $\beta = 2.0$ was obtained by \citet{man2021deposition}. We plot the relationship between $\mathcal{R}$ and $\alpha_{\rm{eff}}$ in Figure \ref{fig:R_alphaeff}, which shows a good collapse of all the simulation data with different mixing ratios and different frictional coefficients. This indicates that, with the assistance from the mixture theory, $\alpha_{\rm{eff}}$ still works for granular systems with two particle species of different frictional properties.

Figure \ref{fig:R_alphaeff} presents a transition from a quasi-static regime to a inertial regime at a transition point at $\alpha_{\rm{eff}}=\alpha_{ce}\approx 3.5$. We regard this transition point as a critical effective aspect ratio, $\alpha_{ce}$. For granular columns with effective aspect ratio less than $\alpha_{ce}$, the final deposition resembles a conical frustum (a truncated cone). During the collapse, the granular column can be divided into an inner static region and an outer flowing region, and the outer flowing region later forms the slope (lateral surface) of the conical frustum. In this case,  the free upper surface of the column also has a static region, which remains to be the upper surface of the resulting conical frustum. As we increase $\alpha_{\rm{eff}}$, the area of the upper surface of the conical frustum decreases, until the conical frustum transforms into a cone, which marks the transition from a quasi-static regime to a inertial regime. In \citet{man2021deposition}, we define a quasi-static column collapse as a system where large portion of the grains remain static, and a inertial column collapse as a system where most grains participate in the flowing region and the inertial effect start to dominate the process. When $\alpha_{\rm{eff}} > \alpha_{ce}$, the entire upper surface of the column starts to flow immediately when we release the particles. 

Thus, the transition point on the $\mathcal{R}(\alpha_{\rm{eff}})$ relationship reflects the position of the surface dividing the static region and the flowing region of the granular column. On one hand, the transition from a quasi-static regime to an inertial regime has its geometric background, since whether forming a conical frustum or forming a cone naturally result in the slope change in the $\mathcal{R}(\alpha)$ relationship. On the other hand, the transition reflects the nature of the yielding condition of the granular system. The static region during a column collapse can be seen as the un-yielded part of the system, while the flowing region is regarded as the yielded part. In previous work, with the analyses of both $\mathcal{R}(\alpha)$ relationship and the $\mathcal{R}(\alpha_{\rm{eff}})$, we have shown that changing either frictional coefficients \citep{man2021deposition} or relative system sizes \citep{man2021finitesize} will affect the transition point. This is similar to the work of \citet{clark2018critical}, showing strong size effects associated to the yielding transition in granular media. However, the analyses of yielding conditions of granular systems is out of scope of this work, we would further analyze this behavior and link the transition in granular column collapses to the yielding criterion.

\subsection{Deposition height}
\label{sec:deposit}
\begin{figure}
  \centerline{\includegraphics[scale=0.32]{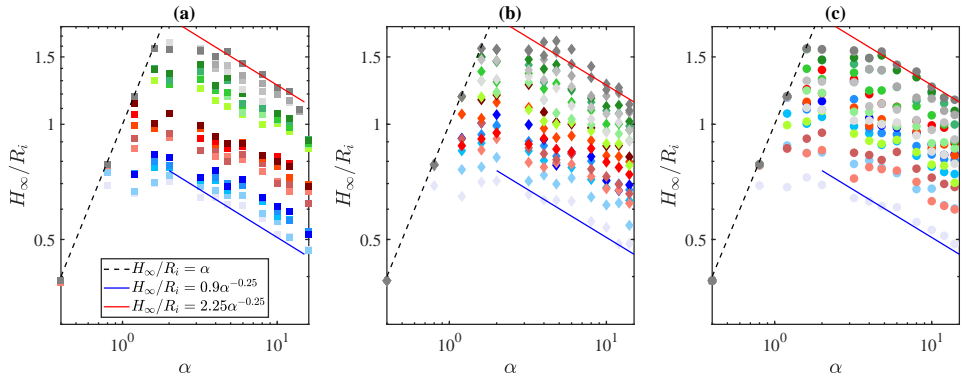}}
  \caption{relationship between the relative deposition height, $H_{\infty}/R_i$, and the initial aspect ratio, $\alpha$, for granular columns with different mixing ratios: (a) Grain\#1 : Grain\#2 = 9 : 1, (b) Grain\#1 : Grain\#2 = 7 : 3, and (c) Grain\#1 : Grain\#2 = 1 : 1. Markers in this figures are the same as those in Figure \ref{fig:R_alpha}.}
\label{fig:height}
\end{figure}

With regard to the deposition height, \citet{lube2004axisymmetric} measured the deposition height, $H_{\infty}$, and plotted $H_{\infty}/R_i$ grainst the initial aspect ratio of granular columns. They observed a collapse of all the experimental data. They conclude that, when $\alpha$ is less than 1.7, $H_{\infty}/R_i$ scales proportionally with repect to $\alpha$, but scales with $\alpha^{1/6}$ when $\alpha$ is larger than 1.7, before $H_{\infty}/R_i$ starts to decrease at $\alpha\approx 6$. However, in our granular system, changing inter-particle friction and mixing ratios dramatically affects the behavior of the deposition height. As shown in Figure \ref{fig:height}, the relationship between $H_{\infty}/R_i$ and $\alpha$ has three distinct parts with two transition points. Different from the work of \citet{lube2004axisymmetric}, the transition points vary with changing frictional coefficients and mixing ratios. To simplify the analysis, we name the transition points as $\alpha_{t1}$ and $\alpha_{t2}$ ($\alpha_{t1} <= \alpha_{t2}$). When $\alpha<\alpha_{t1}$, $H_{\infty}/R_i$ scales proportionally with $\alpha$. When $\alpha\in[\alpha_{t1}, \alpha_{t2}]$, $H_{\infty}/R_i$ almost remain constant. When $\alpha>\alpha_{t2}$, $H_{\infty}/R_i$ starts to decrease as we increase the initial aspect ratio, and approximately $H_{\infty}/R_i \sim \alpha^{-0.25}$, and this often corresponds to a liquid-like regime, as suggested by \citet{man2021deposition}. Similar to our analyses on the run-out distance, the change of mixing ratios influences the spread width of the $H_{\infty}/R_i - \alpha$ relationship, as we keep $\mu_{11}$ constant but vary $\mu_{22}$ from 0.1 to 0.8. For example, in Figure \ref{fig:height}(a) and focusing on maxima of the blue markers (different shades of blue represent different $\mu_{22}$), the maximum $H_{\infty}/R_i$ increases from $\approx 0.7$ to $\approx 0.85$ as we increase $\mu_{22}$. However, when the mixing ratio is $7 : 3$ [shown in Figure \ref{fig:height}(b)], the maxima of blue markers vary from $\approx 0.7$ to $\approx 1.0$. When the mixing ratio is $1 : 1$ [shown in Figure \ref{fig:height}(c)], the maxima of blue markers vary from $\approx 0.7$ to $\approx 1.3$.

The behavior of the $H_{\infty}/R_i - \alpha$ relationship seems reasonable, since increasing frictional coefficient inevitably increases the energy dissipation during the collapse, increases the yielding threshold, and further decreases the relative run-out distance, which results in a larger deposition height. This indicates that the deposition height is also related to the final run-out distance. However, the work of \citet{lube2004axisymmetric}, although works for their experimental data, neglects the influence of $R_{\infty}$ and attributes all the contribution to the initial geometry of the granular column. 

\begin{figure}
  \centerline{\includegraphics[scale=0.4]{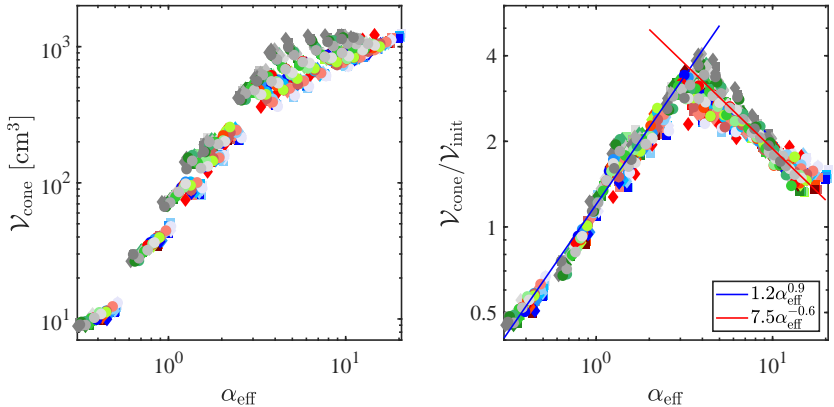}}
  \caption{(a) The relationship between the effective cone volume, $\mathcal{V}_{\rm{cone}}$, and the effective aspect ratio, $\alpha_{\rm{eff}}$, where the effective cone is defined by the deposition height, $H_{\infty}$, and the base radius (deposition radius), $R_{\infty}$. (b) shows the relationship between $\mathcal{V}_{\rm{cone}}/\mathcal{V}_{\rm{init}}$ and $\alpha_{\rm{eff}}$, where $\mathcal{V}_{\rm{init}}$ is the initial radius of the granular column. Markers in this figures are the same as those in Figure \ref{fig:R_alpha}.}
\label{fig:volume}
\end{figure}

In order to consider both the influence of $R_{\infty}$ and frictional properties, we look into the deposition volume, instead of the deposition height, and plot it against $\alpha_{\rm{eff}}$ as shown in Figure \ref{fig:volume}(a). We use the volume of a cone, defined by $H_{\infty}$ and $R_{\infty}$, to represent the deposition situation. Thus, the volume of the deposition cone, $\mathcal{V}_{\rm{cone}}$, is
\begin{equation} \label{eq:coneVol}
    \mathcal{V}_{\rm{cone}} = (\pi/3) R_{\infty}^2H_{\infty}.
\end{equation}
When the resulting deposition of a granular column is a  conical frustum, $\mathcal{V}_{\rm{cone}}$ is usually smaller than the real bulk volume of the collapsed and loosely-packed granular system. Additionally, the difference between the initial solid farction and the final solid fraction may also influence the volume of the deposition cone.

Figure \ref{fig:volume}(a) shows the relationship between $\mathcal{V}_{\rm{cone}}$ and $\alpha_{\rm{eff}}$. When $\alpha_{\rm{eff}} < \alpha_{ce}$, $\mathcal{V}_{\rm{cone}}$ experiences a power-law increase with respect to the increase of $\alpha_{\rm{eff}}$. When $\alpha_{\rm{eff}} > \alpha_{ce}$, the simulation results become scattered. Thus, we cannot obtain a universal relationship between $\mathcal{V}_{\rm{cone}}$ and $\alpha_{\rm{eff}}$, which may indicate that different frictional properties result in different changes of solid fraction before and after the collapse that further affect the deposition height and the resulting conical volume. 

We then calculate the initial volume of the granular column, $\mathcal{V}_{\rm{init}} = \pi R_i^2H_i$, and plot the relationship between $\mathcal{V}_{\rm{cone}} / \mathcal{V}_{\rm{init}}$ against the effective aspect ratio for all the simulation data in Figure \ref{fig:volume}(b). Figure \ref{fig:volume}(b) shows a good collapse of all the simulation data with different frictional coefficients and different mixing ratios. Similar to the $\mathcal{R}(\alpha_{\rm{eff}})$relationship, the relationship between $\mathcal{V}_{\rm{cone}} / \mathcal{V}_{\rm{init}}$ and $\alpha_{\rm{eff}}$ can be divided into two parts, and the transition point locates approximately at $\alpha_{ce}\approx 3.5$. The relationship can be written as
\begin{equation}
  \mathcal{V}_{\rm{cone}} / \mathcal{V}_{\rm{init}} = \left\{
    \begin{array}{ll}
      1.2\alpha_{\rm{eff}}^{0.9}, & \alpha_{\rm{eff}}\le \alpha_{ce}, \\[3.5pt]
      7.5\alpha_{\rm{eff}}^{-0.6}, & \alpha_{\rm{eff}} > \alpha_{ce},
    \end{array} \right.
\end{equation}
which can be used to indirectly calculate the deposition height with the consideration of frictional coefficients and mixing ratios.

\subsection{Kinematic data}
\label{sec:kinematic}
In order to better understand the dynamic behavior of the granular column collapse, we analyzed the data collected of the radius $r$ of the flow front and the average kinetic energy (total kinetic energy of a system divided by the number of particles), and focus on the time when the system reaches the maximum kinetic energy, $t_{\rm{max}}$, and the time when the flow halts, $t_{\rm{f}}$. We have plotted the relationship between the front radius and collapse time and the relationship between the average kinetic energy and the collapse time in Figures \ref{fig:flowBehav10} - \ref{fig:flowBehav50}, which shows that, in the simulations, a granular column collapse can be divided into three stages, which reiterates the results presented in \citet{lube2004axisymmetric}. However, the behavior of $t_{\rm{max}}$ and $t_{\rm{f}}$ shows some differences, which are different from previous works.

\begin{figure}
  \centerline{\includegraphics[scale=0.4]{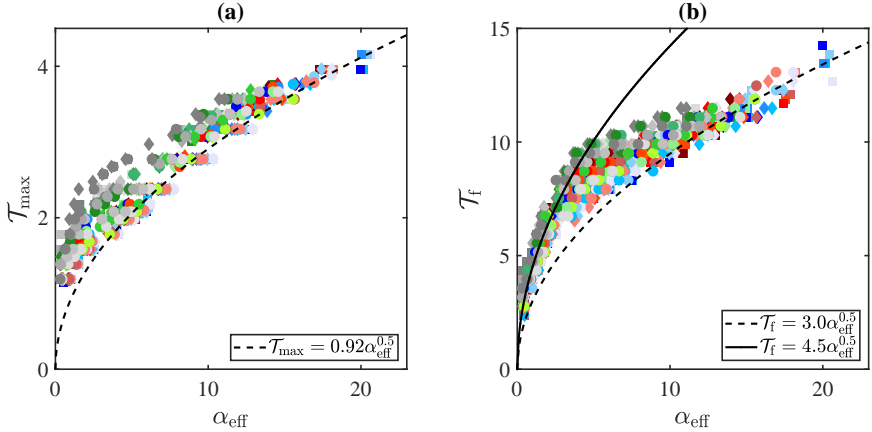}}
  \caption{(a) The relationship between $\mathcal{T}_{\rm{max}} = t_{\rm{max}}/\sqrt{R_i/g}$ and the effective aspect ratio, $\alpha_{\rm{eff}}$, where $t_{\rm{max}}$ is the time when a system reaches its maximum kinetic energy, and $g$ is the gravitational acceleration. (b) The relationship between $\mathcal{T}_{\rm{f}} = t_{\rm{f}}/\sqrt{R_i/g}$ and $\alpha_{\rm{eff}}$, where $t_{\rm{f}}$ is the time when the granular flow halts. Markers in this figures are the same as those in Figure \ref{fig:R_alpha}.}
\label{fig:time}
\end{figure}

We first non-dimensionalize both $t_{\rm{max}}$ and $t_{\rm{f}}$ with respect to a time-scale $\sqrt{R_i/g}$, where $g = 981$ cm/s$^2$ is the gravitational acceleration, so that
\begin{subeqnarray}
    \mathcal{T}_{\rm{max}} &=& t_{\rm{max}}/\sqrt{R_i/g},\\[3pt]
    \mathcal{T}_{\rm{f}} &=& t_{\rm{f}}/\sqrt{R_i/g}.
\end{subeqnarray}
\citet{lube2004axisymmetric} first performed the dimensional analysis and observed a clear scaling law for $\mathcal{T}_{\rm{f}}$, where they fitted $\mathcal{T}_{\rm{f}} \approx 3.0 \alpha^{0.5}$. In terms of the behavior of the kinetic energy, a granular column collapse first experiences a failure process, where potential energy is transformed into kinetic energy, and then experiences a energy dissipation process, where the generated kinetic energy is dissipated by particle collisions. Thus, the time for a system to reach the maximum kinetic energy should scale similarly with the time for a system to stop, which leads to our hypothesis that $\mathcal{T}_{\rm{f}}$ also scales with $\alpha^{0.5}$. Also, since we consider the influence of the inter-particle friction, $\alpha_{\rm{eff}}$, instead of $\alpha$, should be used.

However, different from our hypothesis, the $\mathcal{T}_{\rm{max}} - \alpha_{\rm{eff}}$ relationship, plotted in Figure \ref{fig:time}(a), is scattered. Although its lower bound still scales with $0.92\alpha_{\rm{eff}}^{0.5}$, most simulations experience more time before reaching the maximum kinetic energy. We also observe that decreasing either $\mu_{11}$ or $\mu_{22}$ helps decrease $\mathcal{T}_{\rm{max}}$ and move the $\mathcal{T}_{\rm{max}}(\alpha_{\rm{eff}})$ relationship toward the $0.92\alpha_{\rm{eff}}^{0.5}$ scaling.

We observe similar behavior in the relationship between $\mathcal{T}_{\rm{f}}$ and $\alpha_{\rm{eff}}$, as shown in Figure \ref{fig:time}. When $\alpha_{\rm{eff}} \lessapprox 3.5$, $\mathcal{T}_{\rm{f}}$ follows $4.5\alpha_{\rm{eff}}^{0.5}$. Whereas, when $\alpha_{\rm{eff}} \gtrapprox 13$, $\mathcal{T}_{\rm{f}}$ move back to the classic $3.0\alpha_{\rm{eff}}^{0.5}$ curve. Similar to the relationship between $\mathcal{T}_{\rm{max}}$ and $\alpha_{\rm{eff}}$, systems with higher frictional properties tend to deviate more from the classic scaling curve. Since both $\mathcal{T}_{\rm{max}}$ and $\mathcal{T}_{\rm{f}}$ are associated with the energy generation and dissipation during granular column collapses, we argue that this study, having both larger $\mathcal{T}_{\rm{max}}$ and $\mathcal{T}_{\rm{f}}$, may reflect the initial condition of the granular column, which needs more thorough investigations.

\subsection{Further discussions}
\label{sec:further}
Further analyses are needed to address the deviation of both $\mathcal{T}_{\rm{max}}$ and $\mathcal{T}_{\rm{f}}$ from the classic scaling law. We previously stated that $\mathcal{T}_{\rm{max}}$ and $\mathcal{T}_{\rm{f}}$ can be linked to the energy generation and dissipation during the collapse. Because of the way we generate the Voronoi-based particle packing and the high initial solid fraction $\phi_i = 0.8$, before the column collapse, the granular assembly has a relatively stable structure with usually surface-surface contact among particles, which is different from the situation of spherical particle packing, where the initial solid fraction is often $\phi_i\lessapprox 0.55$ and grains have contact with each other through contact points. This results in more stable granular packing presented in this work, which leads to longer energy generation periods $\mathcal{T}_{\rm{max}}$ than spherical particle packings. The same analysis can be applied to the investigation of $\mathcal{T}_{\rm{f}}$. Thus, we hypothesize that it is the dense and structured initial packing state that results in the longer collapsing period rather than the collapse of a sphere packing.

\begin{figure}
  \centerline{\includegraphics[scale=0.4]{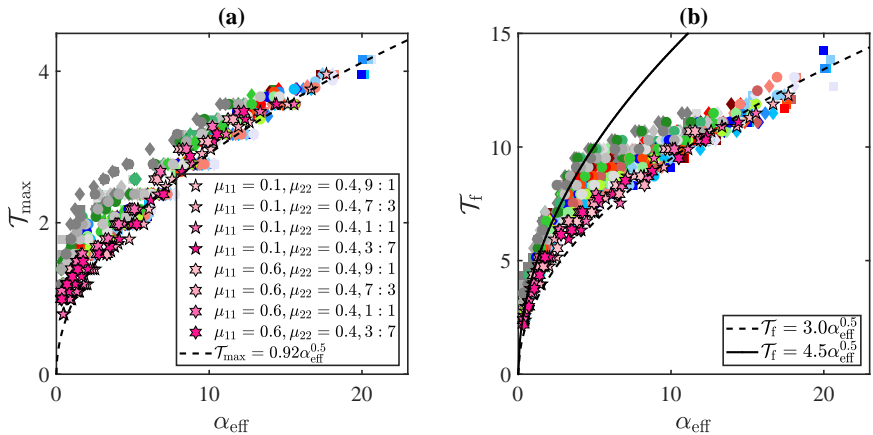}}
  \caption{Same to Figure \ref{fig:time} with additional data from simulations with a low initial solid fraction, $\phi_i = 0.55$. Markers in this figures are the same as those in Figure \ref{fig:R_alpha}.}
\label{fig:time_further}
\end{figure}

In order to weaken the influence of the initial state, we lower the initial solid fraction to $\phi_i = 0.55$ so that the initial granular column is loosely packed. We tested eight sets of additional simulations with (1) $\mu_{11} = 0.1$, $\mu_{22} = 0.4$, mixing ratio $= 9:1$, (2) $\mu_{11} = 0.1$, $\mu_{22} = 0.4$, mixing ratio $= 7:3$, (3) $\mu_{11} = 0.1$, $\mu_{22} = 0.4$, mixing ratio $= 1:1$, (4) $\mu_{11} = 0.1$, $\mu_{22} = 0.4$, mixing ratio $= 3:7$, (5) $\mu_{11} = 0.6$, $\mu_{22} = 0.4$, mixing ratio $= 9:1$, (6) $\mu_{11} = 0.6$, $\mu_{22} = 0.4$, mixing ratio $= 7:3$, (7) $\mu_{11} = 0.6$, $\mu_{22} = 0.4$, mixing ratio $= 1:1$, (8) $\mu_{11} = 0.6$, $\mu_{22} = 0.4$, mixing ratio $= 3:7$. For each set of simulation, we set the initial radius at $R_i = 2.5$ cm and vary its initial height from 1 cm to 35 cm. During the collapse of these granular columns, we measure their $\mathcal{T}_{\rm{max}}$ and $\mathcal{T}_{\rm{f}}$, and plotted both the $\mathcal{T}_{\rm{max}}(\alpha_{\rm{eff}})$ relationship and the $\mathcal{T}_{\rm{f}}(\alpha_{\rm{eff}})$ relationship in Figure \ref{fig:time_further}.

On one hand, Figure \ref{fig:time_further}(a) shows that the additional simulation results, plotted as pentagrams and hexagrams, collapse well onto the scaling curve of $\mathcal{T}_{\rm{max}} = 0.92 \alpha_{\rm{eff}}^{0.5}$, which implies that decreasing the initial solid fraction to a loosely packed level can make the initial packing easier to fail and collapse, which results in a shorter energy-accumulation period (a.k.a, the period for a system to reach the maximum kinetic energy) than the denser granular columns previously simulated. On the other hand, as we plot the results of additional simulations onto the $\mathcal{T}_{\rm{f}}(\alpha_{\rm{eff}}) - \alpha_{\rm{eff}}$ paper, all the simulation data fall nicely onto the classic $\mathcal{T}_{\rm{f}} = 3.0 \alpha_{\rm{eff}}^{0.5}$ curve. We can see that decreasing the initial solid fraction of a granular column dramatically decreases the collapse time. For instance, in the original simulations, for a system with large frictional coefficient and $\alpha_{\rm{eff}} \approx 4$, decreasing $\phi_i$ from 0.8 to 0.55 helps decrease $\mathcal{T}_{\rm{f}}$ from $\approx 10$ to $\approx 7$.

\begin{figure}
  \centerline{\includegraphics[scale=0.4]{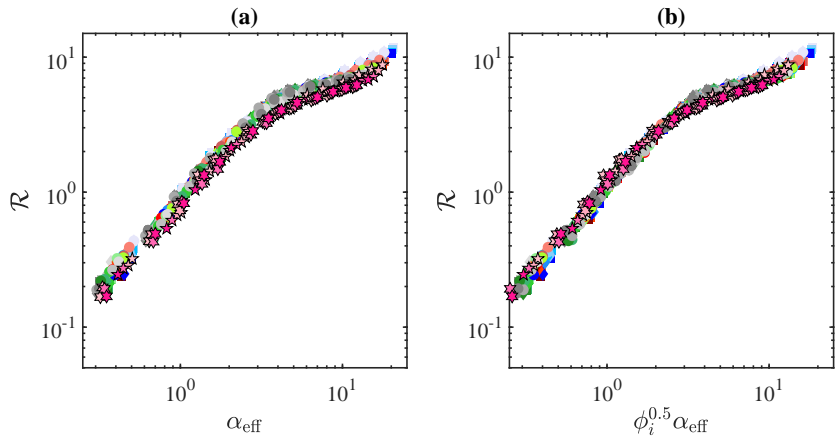}}
  \caption{(a) The relationship between $\mathcal{R}$ and $\alpha_{\rm{eff}}$ with additional data from simulations with a low initial solid fraction, $\phi_i = 0.55$. (b) The relationship between $\mathcal{R}$ and $\phi_i\alpha_{\rm{eff}}$. Markers in this figures are the same as those in Figure \ref{fig:R_alpha}.}
\label{fig:Rfurther}
\end{figure}

We further calculate the relative run-out distance, $\mathcal{R}$, of new simulations, and plot the relationship between $\mathcal{R}$ and $\alpha_{\rm{eff}}$ together with original simulation results in Figure \ref{fig:Rfurther}(a). The new simulation results with $\phi_i = 0.55$ (pentagrams and hexagrams) do not deviate much from the original data, and the $\mathcal{R}$ of systems with $\phi_i = 0.55$ is generally less than that of the original data. We recall the dimensional analysis in \citet{man2021deposition}, where the original form of the effective aspect ratio, $\alpha_{\rm{eff,o}}$, follows
\begin{equation}
    \alpha_{\rm{eff,o}} = \sqrt{\frac{f(\phi_i)}{\mu_g}}\left(\frac{H_i}{R_i} \right)
\end{equation}
where $f(\phi_i)$ is an unknown function of the initial solid fraction and $\mu_g$ is a general form of friction effect, which includes the influence of particle/boundary friction and inter-particle friction. This indicates that change of the initial solid fraction will affect the relative run-out distance. However, in the previous study, since we never changed the initial solid fraction, the influence of $\phi_i$ was neglected and $f(\phi_i)$ was treated as a constant. In this research, as shown in Figures \ref{fig:time_further} and \ref{fig:Rfurther}, changing initial solid fraction lead to different behavior of granular column collapses. We take $f(\phi_i) = \phi_i$ as a simple trial, and plot the relationship between $\mathcal{R}$ and $\phi_i^{0.5}\alpha_{\rm{eff}}$ in Figure \ref{fig:Rfurther}(b), where we still use Eqn. \ref{eq-alpha-eff} as the definition of $\alpha_{\rm{eff}}$. The factor of $\phi_i^{0.5}$ for the $x-$axis helps move both the pentagrams and hexagrams left-ward. The surprisingly good outcome from the fitted $f(\phi_i) = \phi_i$ relationship provide us with a possible option for future studies to include the influence of initial solid fractions. However, the trial of $f(\phi_i) = \phi_i$ is still fitting and without a clear physical interpretation. We will further explore such a influence of $\phi_i$ in future studies.

\section{Conclusions}
\label{sec:conclu}

In this work, we explore the influence of inter-particle frictional coefficients on the axisymmetric collapse of granular columns. The unique aspect of this study is that the granular assembly consists of two species of grains with different inter-particle frictional coefficients. Under such condition of a bi-frictional granular mixture, we have to explore how different mixing ratios influence the final deposition behavior. Three different mixing ratios are considered, where Grain\#1 : Grain\#2 is 9 : 1, 7 : 3, or 1 : 1. In other words, Grain\#2 accounts for 10\%, 30\%, or 50\% of all the particles in a granular column. The collapse of granular columns with different mixtures of frictional coefficients and different initial heights is simulated with DEM with Voronoi-based particles, and we ensure that the bi-frictional granular system is initially well-mixed. We show that the evolution of both the front propagation and average kinetic energy behave the same as granular systems with only one type of particle. The $\mathcal{R}(\alpha_{\rm{eff}})$ relationship remains the same, as we make some modification to the calculation of $\mu_g = \mu_w + \beta\mu_p$, where $\mu_p$ is calculated as the summation of the frictional coefficient of different contact types multiplied by its contact occurrence probability. With the assistance of a simple mixture theory, we can calculate the contact probability of all three different contacts: (1) Grain\#1-Grain\#1 contact, (2) Grain\#2-Grain\#2 contact, and (3) Grain\#1-Grain\#2 contact. The results indicate that the mixing ratio and the contact occurrence probability play an important role in determining the run-out distance of granular column collapses. This might be important when dealing with other problems in granular physics, especially when a granular system has multiple species of grains. 

We show that the analyses of granular column collapses can be further extended to the yielding analysis of granular systems, since the division of the flowing region and the static region is related to the yield criterion, which is influenced by the frictional properties of particles. The frictional coefficient and the mixing ratio influence the percentage of either the flowing region or the static region, which further affects the deposition height of a system. In order to quantitatively describe the deposition height, we introduce a conical volume, $\mathcal{V}_{\rm{cone}}$, calculated from a cone defined by both $H_{\infty}$ and $R_{\infty}$. The resulting relationship between $\mathcal{V}_{\rm{cone}}/\mathcal{V}_{\rm{init}}$ and $\alpha_{\rm{eff}}$ provide us with a method to calculate the deposition height. The relationship between $\mathcal{V}_{\rm{cone}}/\mathcal{V}_{\rm{init}}$ and $\alpha_{\rm{eff}}$ also shows a turning point at $\alpha_{\rm{eff}}\approx\alpha_{ce}$, which is similar to the slope-changing point in the $\mathcal{R}(\alpha_{\rm{eff}})$ relationship.

Although both the $\mathcal{R} - \alpha_{\rm{eff}}$ and the $\mathcal{V}_{\rm{cone}}/\mathcal{V}_{\rm{init}} - \alpha_{\rm{eff}}$ relationships show great collapse of all the data, the dimensionless time for a granular column to reach the maximum kinetic energy, $\mathcal{T}_{\rm{max}}$, and the time for a granular column to rest, $\mathcal{T}_{\rm{f}}$ do not scale nicely with $\alpha_{\rm{eff}}$ and deviate from the classic $\alpha^{0.5}$ scaling. We attribute this phenomena to the initial state of a granular column, i.e., the initial solid fraction and the initial contact structure. To weaken the influence of the initial state, we conduct additional simulations with $\phi_i = 0.55$, showing that decreasing the initial solid fraction helps bring both $\mathcal{T}_{\rm{max}}$ and $\mathcal{T}_{\rm{f}}$ ``on track". We further analyze the influence of the initial solid fraction on the run-out behavior of granular column collapses and recall our previously defined effective aspect ratio, $\alpha_{\rm{eff,o}}$, with $f(\phi_i)$ included. We further propose that $f(\phi_i) = \phi_i$, which can collapse all the simulation data, but lacks of clear physical meaning, which should be further investigated in future studies.

This study strengthens our belief that the run-out behavior of granular columns should be linked to the rheological properties and the yield criterion of granular systems, which implies that the rheology and the failure of granular systems with different species of particles can also follow the same mixture theory to construct a mixed constitutive equation. Also, different system sizes and the corresponding finite-size analysis should be later included in the analyses so that the scaling law could become more physics-based. In fact, the body of the study we have introduced here offers clues on the true form of the rheological law governing the behavior of granular assemblies. Further investigations to link the behavior of idealized granular system and realistic geophysical flows  and include granular rheology into the analysis of granular column collapses are still needed, and will be presented in future publications.

\begin{acknowledgements}
\textbf{Acknowledgements}- We acknowledge the financial support from the National Natural Science Foundation of China with project number 12202367 and 12172305. We thank Westlake University and the Westlake High-performance Computing Center for computational and experimental sources and corresponding assistance. T.M. would like to acknowledge the helpful discussions with Prof. K. M. Hill from the University of Minnesota, and are forever grateful to Ms. X. Luo for her taking extra responsibilities to organize and prepare for their wedding while T.M. was busy doing this research and drafting this manuscript.

\textbf{Declaration of Interests}- The authors report no conflict of interest.
\end{acknowledgements}

\bibliographystyle{jfm}
\bibliography{BifrictionCollapses.bib}

\end{document}